# Widefield phototransient imaging for visualizing 3D motion of resonant particles in scattering environments

Matz Liebel[1,*], Franco V. A. Camargo[2,3], Giulio Cerullo[2,3] and Niek F. van Hulst[1,4,*]

[1] ICFO -Institut de Ciencies Fotoniques, The Barcelona Institute of Science and Technology, 08860 Castelldefels, Barcelona, Spain

[2] Istituto di Fotonica e Nanotecnologie-CNR, Piazza L. da Vinci 32, 20133 Milano, Italy

[3] Dipartimento di Fisica, Politecnico di Milano, Piazza L. da Vinci 32, 20133 Milano, Italy

[4] ICREA - Institució Catalana de Recerca i Estudis Avançats, Passeig Lluís Companys 23, 08010 Barcelona, Spain

*Corresponding authors: matz.liebel@icfo.eu; niek.vanhulst@icfo.eu

**Abstract:** Identifying, visualising and ultimately tracking dynamically moving non-fluorescent nanoparticles in the presence of non-specific scattering is a long-standing challenge across the nano- and life-sciences. In this work we demonstrate that our recently developed ultrafast holographic transient (UHT) microscope is ideally suited for meeting this challenge. We show that UHT microscopy allows reliably distinguishing off-resonant, dielectric, from resonant, metallic, nanoparticles, based on the phototransient signal: a pre-requisite for single-particle tracking in scattering environments. We then demonstrate the capability of UHT microscopy to holographically localize in 3D single particles over large volumes of view. Ultimately, we combine the two concepts to simultaneously track several tens of freely diffusing gold nanoparticles, within a 110x110x110 µm volume of view at an integration time of 10 ms per frame, while simultaneously recording their phototransient signals. The combined experimental concepts outlined and validated in this work lay the foundation for background-free 3D single-particle tracking applications or spectroscopy in scattering environments and are immediately applicable to systems as diverse as live cells and tissues or supported heterogeneous catalysts.

Photothermal imaging and spectroscopy are key methodologies for visualising, quantifying and ultimately studying non-fluorescent nano-objects down to the single molecule level[1–7]. Its unrivalled combination of sensitivity and specificity makes photothermal microscopy, in principle, an ideal candidate for single-particle tracking applications. Unfortunately, the traditional lock-in amplifier-based signal detection schemes that are often necessary to ensure sufficient signal-to-noise ratios are difficult to combine with high-speed imaging over large fields of view, even in relatively simple 2D scenarios[8,9]. Recently, we introduced a nonlinear holographic imaging methodology that enables high-speed, lock-in amplifier-like, signal demodulation using comparatively slow cameras operated in a widefield imaging configuration[10]. When combined with femtosecond pulses, this so-called ultrafast holographic transient (UHT) microscope allowed us to widefield-record pump-probe time-delay dependent transient dynamics of a variety of gold nanoparticles (NPs) with diameters ranging from 20 to 200 nm. Further, the holographic aspect of our approach allowed computational image propagation and refocussing.

In this article we show that UHT microscopy is ideally suited for 3D single-particle tracking, even in its most challenging form: the simultaneous observation of many freely diffusing NPs in large 3D volumes of observation[11]. We first demonstrate that it is possible to discriminate resonant, metallic, from off-resonant, dielectric, NPs with comparable scattering signals, using 2D samples with immobilised particles. Next, we demonstrate that UHT signals are independent of the nominal sample-defocus by using static, pseudo-3D samples, in the form of 2D immobilised NPs that we place at precisely defined z-positions with respect to the image plane. Finally, we address dynamically moving systems by performing UHT-based tracking and time-dependent transient spectroscopy of freely diffusing Au NPs in large, 110x110x110 µm, volumes of observation.

Figure 1 introduces the basic concept of UHT, which we only briefly describe here, referring the interested reader to our proof-of-concept work[10] for a detailed discussion of the more technical and spectroscopy-related aspects. Figure 1a shows a schematic of an UHT microscope[10], a multiplexed off-axis holographic imaging system[12] which is modified to allow generating, demodulating and, ultimately, detecting photoinduced signals with ultrafast pulses (Methods). Briefly, a probe beam illuminates the sample which is imaged onto a conventional CMOS camera in either a dark-field or a bright-field configuration. A 2D 0-π phase grating, conjugate with the camera plane, generates two reference waves which interfere at different angles with the signal scattered/transmitted by the sample. The system outlined above is similar to a conventional multiplexed off-axis holographic microscope[13] which allows using arbitrarily broadband pulses, as the grating ensures interference across the entire camera chip[14–16]. To record photoinduced signals, an additional pump pulse illuminates the sample at a variable, but precisely controlled, pump-probe time-delay. Rather than recording separate sample images in the presence (pump$_{ON}$) or absence (pump$_{OFF}$) of the pump pulse, the UHT microscope simultaneously records the pump$_{ON}$ and pump$_{OFF}$ images in a single camera exposure via a multiplexed holographic demodulation scheme[10].

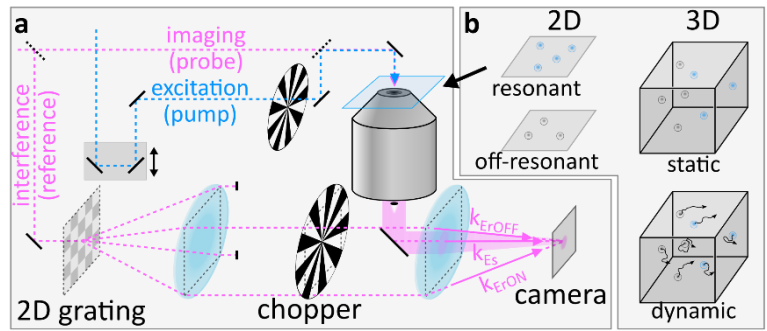

**Figure 1: UHT microscopy applied to a variety of samples.** a) Simplified schematic of an UHT microscope: A multiplexed off-axis holographic microscope is equipped with an additional ultrafast pump laser to selectively photoexcite the sample of interest. b) Summary of typical samples discussed in this work.

The recorded image, $I_{UHT}(x,y)$, can be described as:

$$I_{UHT}(x,y) = E_s^2(x,y) + E_{rON}^2(x,y) + E_{rOFF}^2(x,y) + E_s E_{rON}^*(x,y)e^{-i(ax-by)} + E_s E_{rOFF}^*(x,y)e^{-i(ax+by)} + cc.$$ (Equation 1)

, with (x,y) denoting the camera pixels, $E_s$ the signal field generated by propagating the probe-field through the sample and the imaging system, $Er_{OFF}$ and $Er_{ON}$ the reference waves in the absence and presence of the pump pulse, *cc.* denotes the complex conjugate terms. The camera position dependent linear phase terms, *ax* and *by*, are a direct result of the intentionally non-collinear wave-vectors of $E_s$, $Er_{OFF}$ and $Er_{ON}$ (Figure 1a). Ultimately, these terms enable the well-established Fourier filtering-based approach to off-axis hologram processing[17,18]. As illustrated in Figure 1b, UHT microscopy is very flexible and can be applied to resonant and off-resonant samples, in 2D and 3D, in both static and dynamic configurations.

Figure 2a shows a typical UHT image, acquired for 80 nm Au NPs in dark-field configuration, alongside its Fourier transformation. Interference between the probe and the reference fields directly manifests itself as oscillatory modulation of the NPs' point-spread-functions (PSFs). As mentioned above, a spatial Fourier transformation allows selectively isolating the interference terms $E_s E_{rON}^*$ and $E_s E_{rOFF}^*$ via spatial selection in the Fourier plane followed by removal of the linear phase ramp and an inverse Fourier transformation[10,17,18].

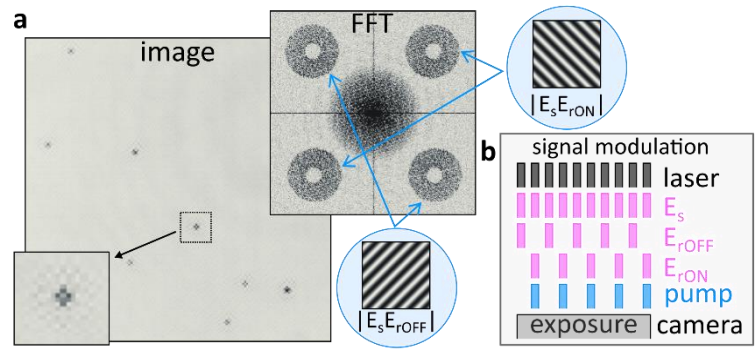

**Figure 2: Image processing and signal demodulation in UHT microscopy.** a) Typical UHT image of multiple 80 nm Au NPs alongside its Fourier transformation and a magnification of a point-spread-function to highlight the interference pattern. The central part of the Fourier transformation (DC) contains the amplitude square terms shown in Equation 1. The four terms along the diagonals are the respective interference terms which are displaced as a result of the linear phase terms. The cartoon insets visualise how linear phase terms would manifest themselves in the most intuitive case of interference between two plane waves. b) UHT modulation scheme that allows encoding and retrieving the desired information from a single camera exposure. Each "bar" represents a laser pulse but the modulation scheme is equally applicable to CW-illumination.

To selectively encode the image-information recorded in the presence (absence) of the pump pulse into $E_s E_{rON}$ ($E_s E_{rOFF}$), we modulate both the pump beam and the reference waves with optical choppers (Figure 2b). The camera detects multiple probe pulses, within a single exposure, which interrogate the sample both with and without pump excitation. The synchronised pump and reference modulation ensures that the $E_{rON}$ ($E_{rOFF}$) reference wave only interferes with probe pulses that interrogate the sample in the presence (absence) of the pump pulse. This all-optical lock-in camera can, in principle, operate at arbitrarily fast demodulation frequencies which are independent of the CMOS camera frame rate, as the demodulation purely relies on computational hologram post-processing and not on fast electronics. As such, the all-optical widefield lock-in modality of the UHT microscope conveniently eliminates the need for point scanning, which is necessary when operating with single-pixel-detector-based lock-in amplifiers, thus making UHT microscopy an ideal candidate for widefield observations of dynamic scenes.

A key asset of UHT microscopy in the context of biological or biomedical imaging is its capability to discriminate resonant particles from the off-resonant scattering background which might arise from cells or tissue[19]. To explore the applicability of UHT microscopy to such scenarios, we compare UHT signals obtained for 60 nm Au and 100 nm latex spheres that we deposit on bare coverglass (Figure 3a). The holographically recorded scattering amplitude of both sets of particles is similar (Figure 3b), thus making them essentially indistinguishable when measured in a linear colour-blind fashion[12]. Their pump-probe time-delay dependent UHT response, however, differs considerably as can be seen by evaluating the scattering signal change computed as the difference between images acquired in the presence and absence of the pump pulse (Figure 3c). The latex NPs show no transient signal, as expected for off-resonantly excited particles probed far off-resonance (Figure 3d). Contrary, the Au

NPs exhibit an instrument-response limited reduction of the differential scattering signal, $\Delta S/S$, due to the generation of a hot-electron distribution[20] followed by thermalisation on the picosecond timescale. We note that the $\Delta S/S$ signal-differences between individual NPs are due to the Gaussian spatial profile of the pump beam, which only weakly photoexcites NPs positioned at its wings.

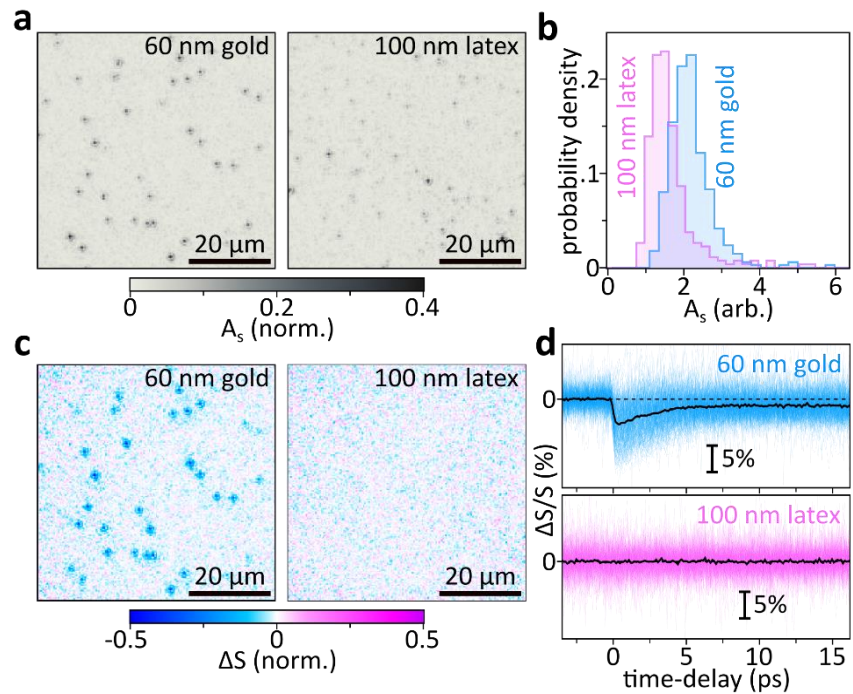

**Figure 3: Transient scattering differentiates resonant and off-resonant particles.** a) Representative holographic scattering images of 60 nm Au and 100 nm latex beads. b) Typical steady-state scattering amplitudes obtained by evaluating the individual scattering signals of 266 latex and 337 gold NPs. c) Normalised scattering signal change for the particles shown in (a). The image is an average over the pump-probe delay window between 0-1.6 ps (15 images averaged). Both images are normalised to the same max/min amplitudes to allow directly comparing the signal magnitudes. d) Pump-probe delay dependent transient scattering signals for all particles evaluated, 337 (gold) and 266 (latex) traces are shown. Pump: 400 nm, probe: 550 nm wavelength.

Beyond being able to simultaneously observe, and identify, many individual nano-objects within a large observation plane, the holographic nature of our method furthermore allows computational image propagation, post-acquisition, to 3D-localise many individual particles from a single camera exposure. Figure 4a demonstrates these capabilities by imaging an out of focus 2D sample composed of many individual 60 nm diameter Au NPs, immobilised on a flat coverglass, at different defocus distances with respect to the image plane. For small defocus distances, similar to those employed in our proof-of-principle work (z = -1040 nm), we still observe many individual particles, which however blur into a speckle-like pattern as we move the sample further away from the image plane (z = 10.7-33.2 µm, Figure 4a, top). Despite this apparent loss of information, computational re-focussing (Methods) of all four recordings results in nearly identical in-focus images (Figure 4a, bottom).

To further quantify these observations, we compare the signals retrieved from the four z-positions. Figure 4b correlates the signal of each particle in all four images with the mean signal of the respective particle. Within our z-range, the scattering signal is essentially independent of the sample's position with respect to the focal plane of the imaging system. Even though this observation might seem surprising at first, the near-identical signal levels are expected if we consider the image-collection and re-focusing operation. Signal loss occurs if the detector fails to record the sample information and it is hence sufficient to employ a detector that is large enough to fully capture the defocused point-spread-functions of the individual particles. Even though defocusing lowers the detected photons per camera pixel, the interferometric detection methodology, with a reference wave that covers the entire detector, ensures camera operation at the shot-noise limit, thus preventing problems associated with entering the dark-noise regime of a camera[21]. Computational re-focussing, a deconvolution operation, will hence always recover the in-focus NP-signal as long as their density is sufficiently low in the field of view and the signal-to-noise ratio is sufficiently high to observe the NP when physically in focus.

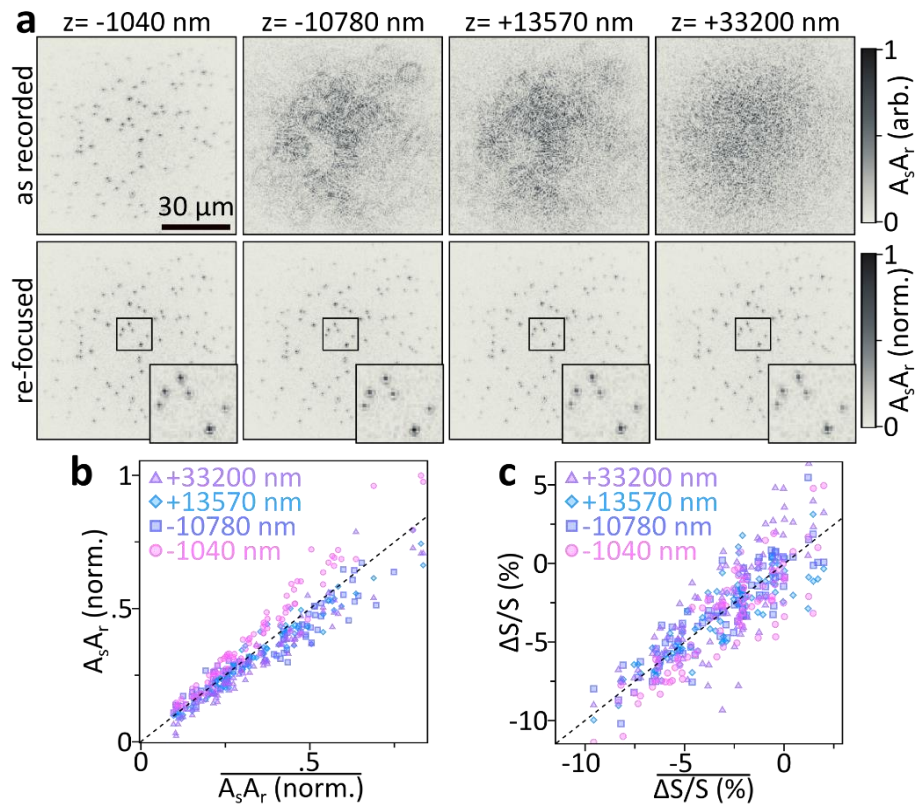

**Figure 4: Out-of-focus holographic and UHT imaging of 60 nm Au NPs.** a) Holographic images recorded at the same xy-sample-position but different z-defocus (top) compared to the images obtained after computational re-focusing (bottom). The re-focused images show nearly identical particle contrast and point-spread-function shape (inset). Only the image obtained after Fourier demodulation is shown. The reconstructed images are normalised to the same maximum value to allow quantitative comparison. b) Correlation between the signal of each individual NP, obtained from the four recorded and re-focused images, and the mean-signal of the same NP. c) As b) but for the mean differential scattering $\Delta S/S$ in the 0-3.6 ps pump-probe time-delay range. Pump: 400 nm, probe: 550 nm wavelength

Thus far, we discussed imaging considerations as applied to 3D single-particle-localisation experiments with linear holography. However, the UHT microscope relies on signals recorded in a pump-probe fashion where an ultrafast pump pulse induces a time-dependent transient signal, or differential scattering ($\Delta S/S(t)$) change. Maximum UHT signals are generally obtained when probing the sample immediately after photoexcitation[10], maximising the magnitude of the photoinduced signal before the onset of relaxation processes. To demonstrate that UHT-based 3D single-particle-localisation experiments are indeed possible, we repeat the re-focusing experiments (Figure 4a,b) but for differential scattering signals measured with the UHT microscope. A 400 nm pump pulse photoexcites the same set of 60 nm Au NPs and a 550 nm probe pulse captures the UHT-response as the average signal in the 0-3.6 ps time-window (computed from 31 individually acquired pump-probe delays). Figure 4c, once again, shows that near-identical signals are obtained for in- and out-of-focus recordings, following digital refocusing. In other words, UHT microscopy should be well suited for 3D nonlinear localisation experiments over large volumes of view.

Following these static proof-of-concept experiments, we now concentrate on freely diffusing particles within a large 3D volume of 110x110x110 µm. Figure 5a shows a 3D representation of typically acquired linear holographic datasets where we combined results of three consecutively performed 3D tracking experiments on a total of 140 freely diffusing Au NPs. To obtain the trajectories shown we first acquire 60 holograms over a total observation time of 30 s, using a camera exposure of 10 ms per image followed by a 500 ms waiting time to maximise the particle motion between frames while reducing the overall data load. We then computationally propagate each hologram over a total of 150 µm, following the same procedure as previously described for the static samples (Figure 4), and then localise all NPs in 3D (Methods). The 3D coordinates of all particles in all 60 frames are then linked to obtain 3D trajectories using the previously followed procedure as outlined in detail by Jaqaman *et al.*[15,16,22].

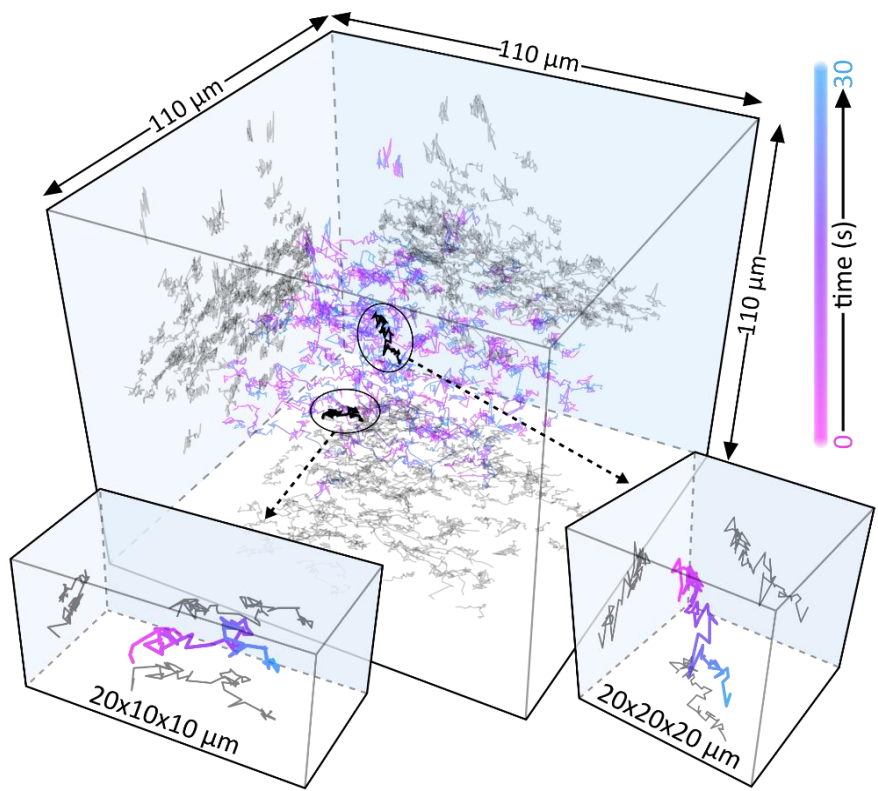

**Figure 5: Large volume-of-view holographic 3D tracking.** a) 3D representation of the trajectories of approximately 140 NPs diffusing in aqueous solution. The pink-to-blue colour coding indicates time, coloured trajectories are 3D, grey trajectories the respective x/y/z projections. Three individually recorded data-sets (each approximately 50 NPs) are combined for representation purposes. The two insets show representative, magnified, particle trajectories.

To demonstrate that the particles observed are, indeed, resonant Au NPs with their corresponding transient dynamics, and hence that 3D tracking can be applied to the transient signals, we follow a different approach. Rather than acquiring the entire video at a fixed pump-probe time-delay, with maximum NP-UHT response, we acquire consecutive images at either negative or positive pump-probe time-delays following the procedure described in Figure 6a. In our approach, we first acquire a single UHT image at positive pump-probe delay, using a 10 ms camera integration time. We then move to a negative delay and record a second image. The two acquisitions are separated by a waiting time of 500 ms to allow sufficient time for physically moving the translation stage (Figure 1a) that controls the pump-probe delay. In this experimental configuration we expect UHT signals at positive time delays and no signal at negative delays, which should allow us to demonstrate that the signal is photoinduced and not a potential artefact due to other experimental factors. The differential scattering histograms reported in Figure 6b show the absence of UHT signals for negative time delays and a marked signal increase at positive delays, thus verifying that UHT microscopy is capable of identifying resonant particles while, simultaneously, following their rapid 3D motion within a large volume of observation. As in the data reported in Figure 4 we, once again, observe a large spread of UHT signals as the pump-beam's fluence is not constant across our large field of illumination. Figure 6c shows examples of time-delay switching on the single-particle level where the individual NPs are centred and re-focused for representation purposes only. One particle is contained in the central area of the pump beam, corresponding to the peak of pump fluence, and the other in its spatial wing, in a region of negligible pump fluence. Even though both particles are visible in the pump$_{OFF}$ and pump$_{ON}$ images, only the NP in the high pump fluence area exhibits a UHT signal and only for positive time-delays.

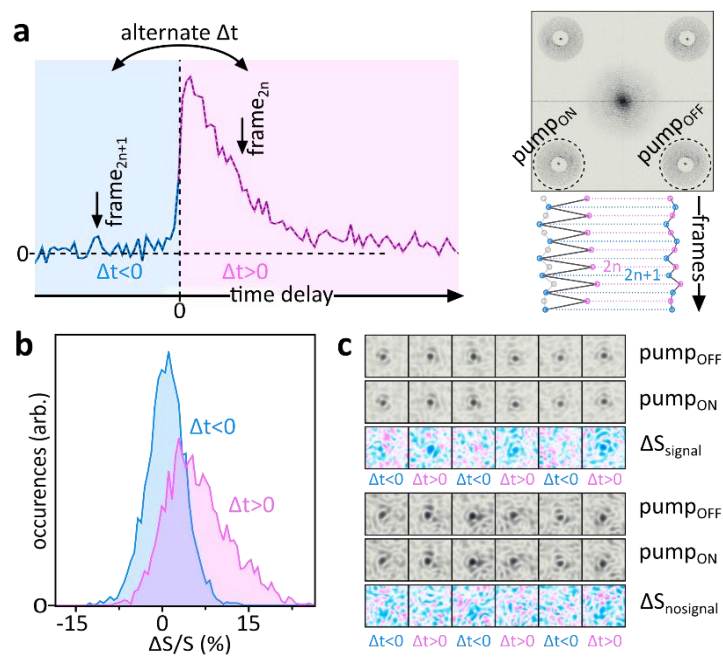

**Figure 6: In-situ modulation of UHT signals for freely diffusing particles.** a) Alternating the pump-probe time-delay between image acquisitions switches the UHT signal on and off (left) which allows identifying resonant NPs as long as the same particle can be followed over several frames (right). b) Comparison between the differential scattering signal histograms of all freely diffusing NPs acquired before (blue) and after (pink) photoexcitation. We note that the $\Delta S/S$ signal inversion, as compared to Figure 4c, is due to the different size of the NPs, a different probe wavelength and the fact that the NPs are immersed in water (n=1.334) rather than immobilized on a glass surface in air. c) Comparison of pump$_{OFF}$ and pump$_{ON}$ images and difference between the two ($\Delta S$) for a particle contained in a high-fluence region (top) or a region of negligible pump-fluence (bottom). The freely diffusing particles are computationally re-focused and centred for representation purposes only, their physical position changes due to Brownian motion. Imaging is performed using a camera exposure of 10 ms followed by a 500 ms waiting-time to allow physical movement of the translation stage; pump: 400 nm, probe: 566 nm wavelength.

We finally perform pump-probe time-dependent transient scattering spectroscopy on a freely diffusing 100 nm Au NP. Figure 7a shows the as-recorded holographic image acquired with the focus placed in the vicinity of the glass-water interface. The two visible particles are stuck to the glass surface. We computationally propagate by approximately 10 µm above the recording plane where we identify a freely diffusing NP (Figure 7b, top). By continuously maximizing its amplitude via computational propagation, we are able to follow the xyz-motion of said particle for holographic images acquired at different times, as shown in Supplementary Video 1. Simultaneously, we record UHT images where the particle is only visible for positive time delays (Figure 7b, bottom), in agreement with the observations made previously (Figure 6). Finally, we use UHT microscopy to follow the 3D trajectory of the NP while changing the pump-probe time delay, which allows recording the 3D trajectory as well as the transient dynamics of a rapidly diffusing object (Figure 7c and Supplementary Video 1). As expected, the $\Delta S/S$ signal of the gold NP decays on the picosecond timescale due to hot electron cooling via coupling with phonons.

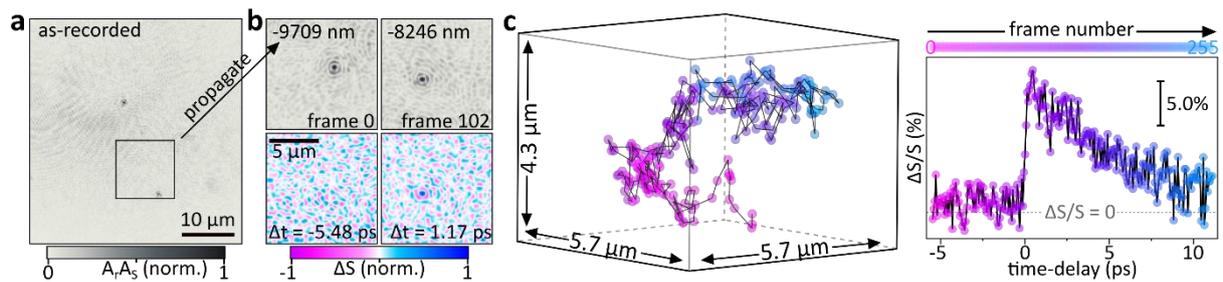

**Figure 7: Transient spectroscopy of freely moving objects.** a) As acquired image recorded at the glass-water interface highlighting the region of interest where a freely diffusing 100 nm Au NP is identified by b) computationally propagating approximately -9709 nm (frame 0) or -8246 nm (frame 102). Transient UHT images obtained at the same frames but for two different pump-probe delays are shown below (see Supplementary Video 1). c) 3D trajectory of the freely diffusing NP shown in (a) alongside the simultaneously recorded transient scattering time-trace. 500 ms waiting-time between frames, 10 ms integration; pump: 400 nm, probe: 566 nm wavelength.

To summarise, we showed that our UHT microscope[10] is ideally suited for single-particle tracking applications over large volumes of view. Analogous to photothermal microscopy, UHT microscopy allows identifying and isolating resonant particles, such as the gold NPs employed in this work, from off-resonant particles, which is the key requirement for imaging applications in scattering environments such as live cells or tissue[9]. The holographic all-optical lock-in modality exploited in UHT microscopy further eliminates all limitations of traditional photothermal imaging, where lock-in amplifier based point-scanning severely limits the technique's applicability to observing dynamic systems, even in a simple 2D context[9]. The relatively slow frame-rates of 2 frames-per-second employed here were solely chosen to allow changing the pump-probe time-delay between image acquisitions, which was necessary to conclusively demonstrate that the UHT signals observed are indeed pump-induced and not an artefact of our imaging system. For high-speed tracking applications the camera exposures of 10 ms, used here, correspond to a nominal frame rate of 100 frames-per-second and it should easily be possible to further increase the frame rate to up to half the laser's repetition rate by employing a faster camera and further boost the sensitivity by using a laser with higher repetition rate. These capabilities make UHT microscopy an ideal candidate for both high-speed and long-term tracking of resonant particles in 3D scattering environments. As such, we envision immediate applications in the context of cellular particle-internalisation dynamics of theranostic or nanomedical agents as well as their transport through cells and tissue.

**Methods:**

**Microscope:** The UHT microscope employed in this work is analogous to our previous implementation[10] based on a 1-kHz, 100-fs amplified Ti:sapphire laser[23] delivering both the 400 nm pump pulse as the second harmonic and the probe pulse via a home-built non-collinear optical parametric amplifier (NOPA). A 90:10 beam-splitter generates the probe and the reference from the NOPA output. A F=200 mm achromatic lens focuses the collinear pump and probe pulses onto the sample. Residual pump light is removed with a 488 nm long-pass filter (*488 nm EdgeBasic™, Semrock*) and the probe scattering is collected with a home-built transmission dark-field microscope (NA=0.5, *Olympus RMS20X-PF*) and imaged onto a CMOS camera (*acA2040-90um Basler ace, Basler AG*) at a magnification of 37x. Off-axis holography is implemented by interfering the sample scattering with two of the first diffraction orders of a 2D 0-π phase grating (25.6 grooves/mm) which we relay-image (nominal magnification 0.5x) onto the same CMOS camera. Two synchronized mechanical choppers are used to modulate both the reference waves as well as the pump pulse as described in detail previously[10] and in the main text. The signal to reference time-delay is adjusted with a mechanical delay line. Pump-probe time-delay dependent images are recorded in an automatized fashion with a computer-controlled translation stage (*M-531.PD1, Physik Instrumente*).

**Sample preparation:** #1.5 coverglass is cleaned by 10 min sonication in acetone, 10 min sonication in MilliQ, drying under a stream of $N_2$ followed by 5 min oxygen plasma treatment (*Femto, Diener electronic GmbH*). The cleaned glass is then incubated with PLL-g-PEG (*SuSoS AG*), for approximately 10 min, followed by a dilute 60 nm Au NP-solution (*BBI Solutions*). The latex NP-sample is fabricated by drying 5 µl of 100 nm polystyrene particles (*Sigma Aldrich*) on glass. For experiments on freely diffusing particles we dilute the NP stock solution using MiliQ water containing 0.5% polyacrylic acid to reduce the NPs' mobility. The solutions are imaged using a coverglass sandwich with separation between the two glasses of ≈200 µm.

**Experimental parameters:** Due to the Gaussian beam-shapes the fluences vary widely across the field of observation and we solely quote 1/e values for Figures 4,5 with nominal fluences being: 1.00 mJ/cm² pump, 0.75 mJ/cm² probe, 10 ms integration time.

**Angular spectrum method:** We perform image-propagation via the angular spectrum method[24]. Briefly the processed *NxN* holograms are convolved with a propagation kernel of the form:

$$K(x, y, z) = \exp(iz\sqrt{k_m^2 - k_x^2 - k_y^2}),$$

where $k_m = 2n\pi/\lambda$, with *n=1* being the refractive index of air. The discretized spatial frequencies are $(k_x, k_y) = 2\pi/n\Delta x(x,y)$ for $(-N/2 \leq x, y < N/2)$, with *Δx* representing the magnified pixel size of the imaging system.

**3D localization, large volume:** For 3D localization, each hologram is propagated from –75 µm to +75 µm with a spacing between different z-planes (*dz*) of 500 nm. The resulting image cube is then used to identify candidate particles by applying multiple convolutions in combination with noise and particle-size based thresholding[25]. To achieve sub-pixel localization in the *x,y*-coordinates, particles that are in focus at a specific *z*-plane are fitted by a 2D Gaussian. For the z-coordinate, sub-*dz* localization is achieved by first calculating the Tamura values ($T(z) = \sqrt{\sigma(lz)/mean(lz)}$) for a region of interest of (≈ 3x3 µm), centred about the intensity maxima for each z-plane, and then fitting a parabola using the two most adjacent pixel values along the maximum[15].

**Acknowledgments**

Authors acknowledge support by the Spanish Ministry of Science, Innovation and Universities (MCIU/AEI: RTI2018-099957-J-I00 and PGC2018-096875-B-I00), the Ministry of Science and Innovations (MICINN "Severo Ochoa" program for Centers of Excellence in R&D CEX2019-000910-S), the Catalan AGAUR (2017SGR1369), Fundació Privada Cellex, Fundació Privada Mir-Puig, and the Generalitat de Catalunya through the CERCA program. N.F.v.H. acknowledges the financial support by the European Commission (ERC Advanced Grant 670949-LightNet). G.C. acknowledges support by the European Union Horizon 2020 Programme under Grant Agreement No. 881603 Graphene Core 3.